\documentclass[12pt]{article}
\usepackage{times}
\usepackage{geometry}
\usepackage[utf8]{inputenc}
\usepackage{enumitem,amssymb}
\usepackage{ragged2e}
\newlist{thematic}{itemize}{8}
\setlist[thematic]{label=$\square$}
\usepackage{pifont}
\usepackage{etoolbox}
\usepackage{xspace}
\usepackage{xcolor}
\usepackage{hyperref}
\hypersetup{
    colorlinks=true,
    linkcolor=blue,
    filecolor=magenta,      
    urlcolor=cyan,
}
\usepackage{graphicx}
\graphicspath{ {./figures/} }

\newtoggle{long_names}
\togglefalse{long_names}
\toggletrue{long_names}      % remove comment to get longer institution names

\iftoggle{long_names}{
  \def\cu{ (Cornell University)}
\def\ubon{ (University of Bonn)}
\def\ucol{ (University of Cologne)}
\def\mpa{ (Max-Planck-Institute for Astrophysics)}
\def\ucam{ (University of Cambridge)}
\def\uwat{ (University of Waterloo)}
\def\cita{ (Canadian Institute for Theoretical Astrophysics)}
\def\usta{ (Stanford University)}
\def\umic{ (University of Michigan)}
\def\dalu{ (Dalhousie University)}

\def\nist{ (National Institute of Standards and Technology)}
 \def\smu{ (Southern Methodist University)}
 \def\asu{ (Arizona State University)}
 
 \def\puc{ (Pontificia Universidad Cat\'olica de Chile)}
 \def\nrc{ (NRC Herzberg Institute of Astrophysics)}
\def\ucsc{ (Universidad Cat\'{o}lica de la Santísima Concepci\'{o}n)}
\def\colo{ (University of Colorado, Boulder)}
\def\udp{ (Universidad Diego Portales)}
\def\queens{ (Queen's University)}
\def\abert{ (University of Alberta)}
\def\utor{ (University of Toronto)}
\def\rug{ (Rijks Universiteit Groningen)}
}{
  \def\cu{ (Cornell)}
\def\ubon{ (Bonn)}
\def\ucol{ (Cologne)}
\def\mpa{ (MPA)}
\def\ucam{ (Cambridge)}
\def\uwat{ (Waterloo)}
\def\cita{ (CITA)}
\def\usta{ (Stanford)}
\def\umic{ (Michigan)}
\def\dalu{ (Dalhousie)}

\def\nist{ (NIST)}
 \def\smu{ (Southern Methodist)}
 \def\asu{ (ASU)}
 
\def\puc{ (Cat\'olica)}
\def\nrc{ (NRC-Herzberg)}
\def\ucsc{ (UCSC)}
\def\colo{ (CU-Boulder)}
\def\udp{ (UDP)}
\def\queens{ (Queen's)}
\def\abert{ (Alberta)}
\def\utor{ (Toronto)}
\def\rug{ (RUG)}
}

\newcommand  \gtsim  {\lower.5ex\hbox{$\; \buildrel > \over \sim \;$}}
\newcommand  \ltsim  {\lower.5ex\hbox{$\; \buildrel < \over \sim \;$}}

\newcommand	\micron	{{$\mu${\rm m}}}
\newcommand{\um}{\micron\xspace}

\def\3he{$^3{\rm He}$}

\def\pcam{Prime-Cam\xspace}
\def\ccatp{CCAT-prime\xspace}

\def\deg{$^{\circ}$}

\begin{document}
\raggedright
\huge
Astro2020 APC White Paper \linebreak
\vspace{-0.05in}

The CCAT-Prime Submillimeter Observatory \linebreak
\normalsize
\vspace{0.05in}

\noindent \textbf{Thematic Areas:} \hspace*{61pt} 
$\square$ Planetary Systems \hspace*{9pt} 
\fbox{X} Star and Planet Formation
\linebreak
$\square$ Formation and Evolution of Compact Objects \hspace*{31pt} 
\fbox{X} Cosmology and Fundamental Physics 
\linebreak
\fbox{X} Stars and Stellar Evolution \hspace*{6pt} 
$\square$ Resolved Stellar Populations and their Environments 
\linebreak
\fbox{X} Galaxy Evolution   \hspace*{44pt} 
$\square$ Multi-Messenger Astronomy and Astrophysics 
\linebreak
  
\noindent \textbf{Project Director:} Terry Herter$^*$ (Cornell University)
\linebreak
\vspace{-0.1in}

\justifying

\noindent  \textbf{Co-authors:} Manuel Aravena\udp, Jason Austermann\nist, Kaustuv Basu\ubon, Nicholas Battaglia\cu,  Benjamin Beringue\ucam, Frank Bertoldi\ubon, J. Richard Bond\cita, Patrick Breysse\cita, Ricardo Bustos\ucsc, Scott Chapman\dalu, Steve Choi\cu, Dongwoo Chung\usta, Nicholas Cothard\cu,  Bradley Dober\colo, Cody Duell\cu, Shannon Duff\nist, Rolando D\"unner\puc, Jens Erler\ubon, Michel Fich\uwat, Laura Fissel\queens, Simon Foreman\cita, Patricio Gallardo\cu, Jiansong Gao\nist, Riccardo Giovanelli\cu, Urs Graf\ucol, Martha Haynes\cu, Terry Herter\cu, Gene Hilton\nist, Ren\'{e}e Hlo\v{z}ek\utor, Johannes Hubmayr\nist, Doug Johnstone\nrc, Laura Keating\cita, Eiichiro Komatsu\mpa, Benjamin Magnelli\ubon, Phil Mauskopf\asu, Jeffrey McMahon\umic, P. Daniel Meerburg\rug, Joel Meyers\smu, Norm Murray\cita, Michael Niemack$^*$\cu, Thomas Nikola\cu, Michael Nolta\cita, Stephen Parshley\cu, Roberto Puddu\puc, Dominik Riechers\cu, Erik Rosolowsky\abert, Sara Simon\umic, Gordon Stacey$^*$\cu, Jason Stevens\cu, Juergen Stutzki\ucol, Alexander Van Engelen\asu, Eve Vavagiakis\cu, Marco Viero\usta, Michael Vissers\nist, Samantha Walker\colo, Bugao Zou\cu

\vspace{0.1in}

\noindent $^*$ Corresponding authors: 

Terry Herter, tlh10@cornell.edu

Michael Niemack, niemack@cornell.edu

Gordon Stacey, stacey@astro.cornell.edu

\pagebreak
\medskip

\noindent \textbf{Executive Summary: }
% going to bulletized form - this may be all that gets read!
The Cerro Chajnantor Atacama Telescope-prime (\ccatp) is a new 6-m, off-axis, low-emissivity, large field-of-view submillimeter telescope scheduled for first light in the last quarter of 2021.  In summary,
\vspace{-0.05in}
\begin{enumerate}
\itemsep 0em
\item[(a)] \ccatp uniquely combines a large field-of-view (up to 8-deg), low emissivity telescope ($<$ 2\%) and excellent atmospheric transmission (5600-m site) to achieve unprecedented survey capability in the submillimeter.
\item[(b)] Over five years, \ccatp \textit{first generation science} will 
address the physics of star formation, galaxy evolution, and galaxy cluster formation; probe the re-ionization of the Universe; improve constraints on new particle species; and provide for improved removal of dust foregrounds to aid the search for primordial gravitational waves.
\item[(c)] The Observatory is being built with non-federal funds ($\sim$ \$40M in private and international investments).  Public funding is needed for instrumentation ($\sim$ \$8M) and operations
(\$1-2M/yr). In return, the community will be able to participate in survey planning and gain access to curated data sets.
\item[(d)] For \textit{second generation science}, CCAT-prime will be uniquely positioned to contribute high-frequency capabilities to the next generation of CMB surveys in partnership with the CMB-S4 and/or the Simons Observatory projects or revolutionize wide-field, sub-millimetter line intensity mapping surveys. 
\end{enumerate}

\setlength{\parskip}{0.75em}

\newpage
\setcounter{page}{1}

\begin{figure}[t]
\begin{center}
\includegraphics[width=6.5in]{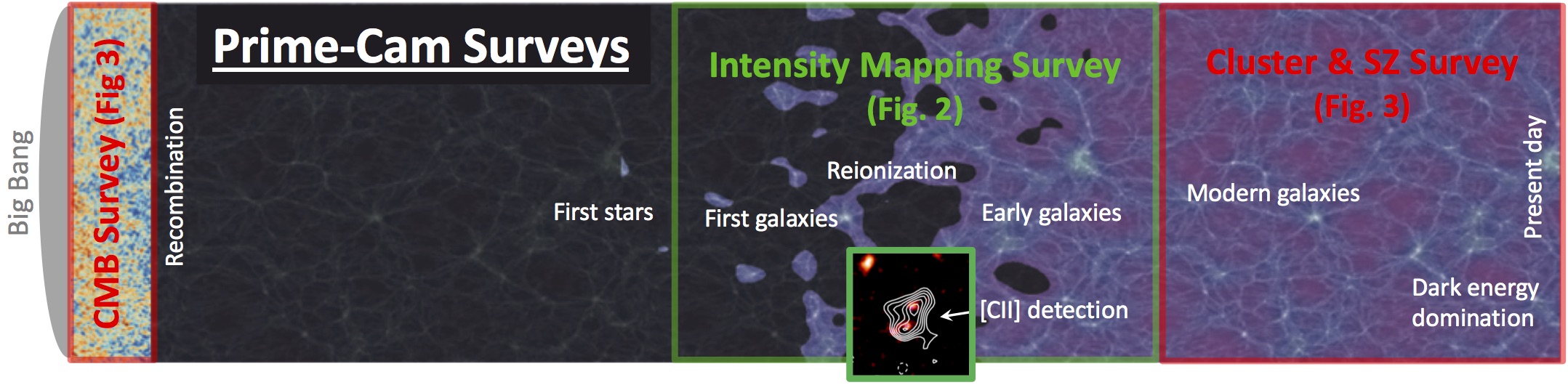}
\end{center}
\vspace{-4mm}
\caption{
\small  \pcam surveys on a schematic history of the Universe. The three primary science surveys in key legacy survey fields targeted by the planned \pcam observations are: [CII] line intensity mapping of the Epoch of Reionization (EoR, Figure~\ref{fig:EoR_IM}), galaxy cluster and Sunyaev-Zeldovich (SZ, Figure~\ref{fig:SZ}), and CMB polarization (CMB, Figure~\ref{fig:SZ}; image modified from \cite{robertson10}; 
[CII] detection inset from \cite{riechers14b}). 
}
\label{fig:cmb_to_now}
\end{figure}

\textbf{\Large Key Science Goals and Objectives}

\textbf{\it \large First generation science:}

During its first five years, \ccatp will perform a set of wide-area surveys to uniquely address forefront science. Below we give brief descriptions of the different areas.  References to relevant Astro2020 White Papers follow the science theme titles. 

% EoR IM science

\textbf{Tracing the epoch of reionization with CII intensity mapping} \cite{kovetz19,chang19,cooray19, astro2020Meerburg}:  \ccatp will aid in understanding the processes of structure formation and probe the underlying dark matter density fluctuations by  mapping the large-scale three-dimensional (3D) clustering of star-forming galaxies.  This is done using wide-field spectroscopy of the aggregate [CII] 158\um signal from star formation regions in galaxies to 
perform 3D tomography of the 
spatial fluctuations due to large-scale structure at $z$ = 3.5 - 8.0, all the way from ``cosmic noon'' into the epoch of reionization \cite{kovetz17}.  

The Prime-Cam instrument \cite{stacey/etal:2018,vavagiakis/etal:2018}, described below in the technical overview, will provide maps over several 8\,deg$^2$ survey fields in [CII] emission across this redshift range (Figure 2). 
It will simultaneously provide measurements of the [OIII] 88\um line from HII regions in $z>7$ galaxies and of multiple CO lines at virtually any redshift. This data will be highly complementary to targeted studies of individual galaxies at the same redshifts with ALMA, ngVLA, JWST, and OST by providing information on the large-scale structure in which those sources are embedded and the aggregate signal from sources too faint to be individually detected by those facilities.

\begin{figure}[t]
\begin{center}
\includegraphics[width=6.0in]{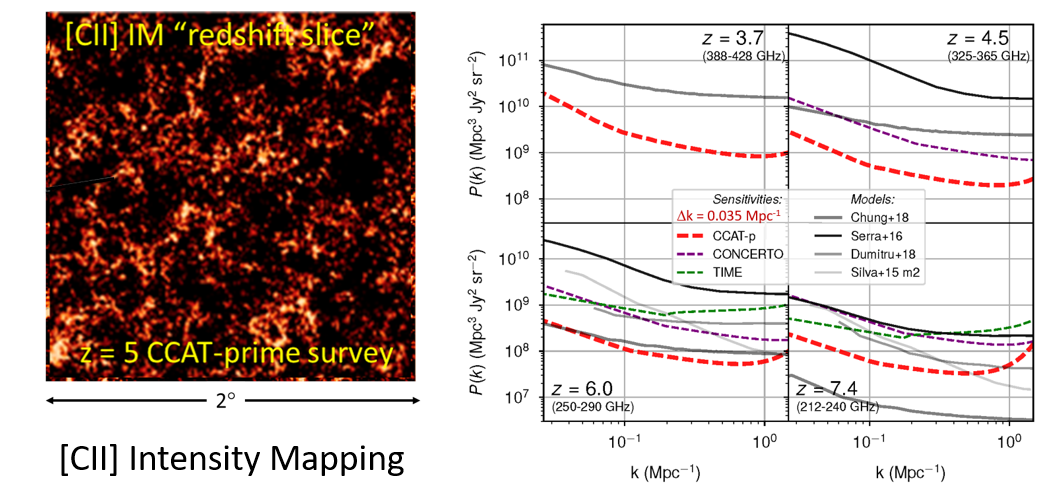}
\end{center}
\vspace{-4mm}
\caption{\small 
%{\bf EoR Survey ---}  
Simulated ``redshift slice'' ({\it left}) of the area covered by the [CII] intensity mapping survey in the epoch of reionization (EoR), representing a small spectral bin within the vast $z=3.5$--8.0 redshift range covered by the \pcam spectrometer [CII] measurements. {\it Right:}  The power spectrum of the [CII] emission reveals the topology of reionization and the [CII] luminosity fluctuations at redshifts of 3.7 to 7.4 
(predictions based on \cite{chung/etal:2018}).  Recent state-of-the-art model predictions (solid lines) differ by factors of 10 to 50. Overlaid are our predicted EoR-Spec sensitivities for the planned 16\,deg$^2$ first generation survey.  
\pcam surveys are as much as 10$\times$ more sensitive than other proposed surveys (TIME and CONCERTO; \cite{Crites:2014,lagache18}) 
leading to an expectation for detection of the intensity mapping signal at redshifts approaching 8 for most models. 
ALMA cannot make these measurements, since its field of view is smaller than a single pixel of EoR-Spec on CCAT-prime, and because it is mainly sensitive to the signal of individual galaxies, rather than the aggregate emission seen by intensity mapping studies. 
}
\label{fig:EoR_IM}
\end{figure}

% SZ science

\textbf{Understanding galaxies and galaxy cluster formation via millimeter and submillimeter observations} \cite{astro2202Basu,astro2020Mantz,astro2020Battaglia,mroczkowski19}: \ccatp will constrain feedback mechanisms and test cosmological simulations by measuring the thermodynamic properties of galaxies and galaxy clusters via the Sunyaev-Zel'dovich (SZ) effects on the cosmic microwave background (CMB) \cite{mittal/etal:2017,Battaglia17}. \ccatp's 220--410\,GHz measurements combined with 30--270\,GHz data from Advanced ACTPol (AdvACT) \cite{henderson/etal:2016} and Simons Observatory (SO) \cite{galitzki/etal:2018} will measure $\sim$ 16,000 clusters and detect the fainter kinetic SZ (kSZ) effect to trace the peculiar velocity field in the Universe (the standard thermal SZ effect, tSZ, will be readily detected). The data will also allow for precise tracing of the SZ spectral shapes of individual clusters enabling measurements of their mean electron temperature (the relativistic SZ effect, rSZ).  

With these measurements \ccatp will provide valuable complementary information for deriving accurate cluster properties, such as masses. \pcam measurements have the potential to enable the detection of all three contributors to SZ components (cluster optical depth, bulk velocity, and temperature) and dust emission from individual clusters \cite{Erler2018, mittal/etal:2017} for a large, statistically significant cluster sample for the first time. Measurements of the tSZ and kSZ effects provide windows into the thermodynamic profiles of individual galaxy clusters and ensemble samples of galaxies. These will enable essential tests for feedback mechanisms and for the successful cosmological hydrodynamical simulations that match the optical properties of galaxies in dense environments measured by space- and ground-based optical surveys, such as LSST.

\begin{figure}[t]
\begin{center}
\includegraphics[width=6.2in]{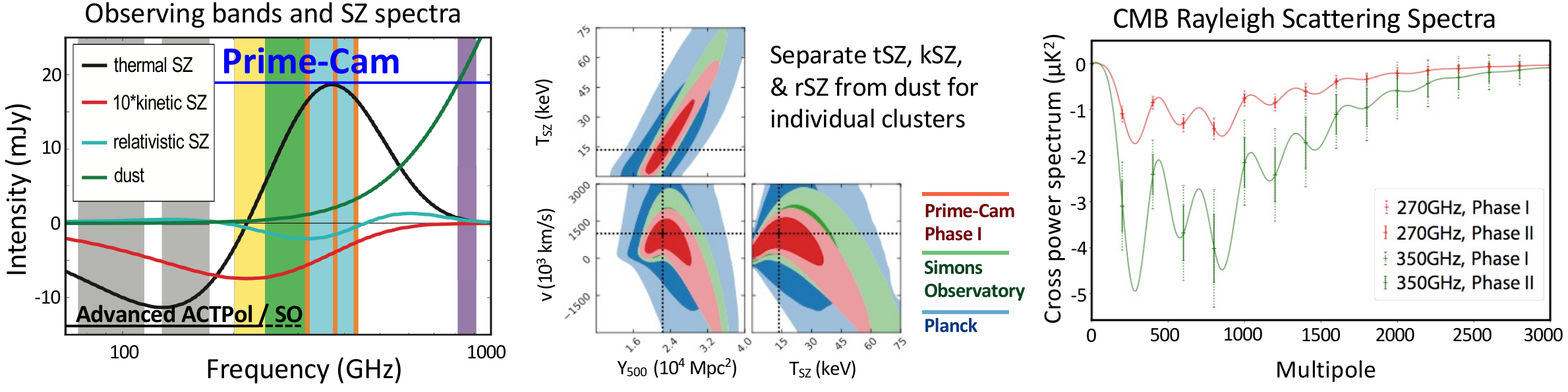}
\end{center}
\vspace{-4mm}
\caption{\small 
\pcam observations will span from 200 to 900\,GHz ({\it left}), including photometric (green, blue, and purple bands) and spectroscopic observations (yellow through orange bands). Measurements at these frequencies are needed in addition to lower frequency bands
(grey and yellow - AdvACT - plus green - SO) to cleanly separate CMB and SZ signals (black, red, blue lines) from dust (green line) that contaminates CMB anisotropies. 
These observations will enable determination of the temperatures and velocities of galaxy clusters ({\it middle}, $1\sigma$ and $2\sigma$ contours shown) \cite{Erler2018}.  
High frequency CMB polarization measurements with \pcam will enable high-S/N measurements of Rayleigh scattering of the CMB for the first time ({\it right}), which will lead to improvements in cosmological constraints beyond what can be achieved with traditional lower-frequency 
CMB measurements \cite{Alipour:2014dza}. 
Such improvements in cosmological constraints could provide evidence for beyond the standard model particle species and possibly dark matter candidates \cite{BeringueMeerburgMeyersinprep}. 
}
\label{fig:SZ}
\end{figure}

\textbf{Enhancing constraints on new particle species with Rayleigh Scattering \cite{astro2020Green}:} \ccatp will measure Rayleigh scattering of CMB photons for the first time to improve constraints on new particle species 
\cite{abazajian/etal:2016,Alipour:2014dza}. \pcam's 270--350\,GHz bands open up yet another unexplored window of Rayleigh scattering. Shortly after recombination, CMB photons undergo Rayleigh scattering with neutral hydrogen and helium with a cross-section that is strongly frequency-dependent \cite{Lewis:2013yia}. This has the effect of a frequency-dependent change to the visibility function, suppressing primary CMB temperature fluctuations and generating new polarization at high frequencies and small angular scales, providing access to new modes that probe the early Universe \cite{Lewis:2013yia} and improvements on cosmological parameters affecting recombination physics \cite{Alipour:2014dza}.

\ccatp is uniquely positioned to make the first detection of Rayleigh scattering of the CMB\footnote{It may be possible to detect Rayleigh scattering in the Planck CMB maps. However, it is expected to be at lower significance and \ccatp's wide and high resolution frequency coverage will help mitigate foregrounds more effectively \cite{BeringueMeerburgMeyersinprep}.} (Fig.~\ref{fig:SZ}) and utilize the information from it in future cosmological constraints.  The forecasted improvement in cosmological parameters is modest, however, without Rayleigh scattering the constraints on beyond the standard model (BSM) particle species and possible dark matter candidates requires a factor of $>2$ increase in number of detectors or integration time. Thus, \ccatp could improve the science return and/or reduce costs by tens of millions of dollars for future,  ground-based CMB experiments (CMB-S4 and Simons Observatory) that aim to further constrain these BSM particle species.

\textbf{Measuring CMB foregrounds to aid the search for primordial gravitational waves \cite{astro2020Kallosh}:} \ccatp will aid the search for primordial gravitational waves that may have been generated during an inflationary epoch by characterizing dust foregrounds that dilute measurements of CMB polarization \cite{abazajian/etal:2016}. By measuring dust polarization properties above 250 GHz with sub-arcmin resolution and high signal-to-noise ratios using \pcam, we will be able to measure the spectrum and polarization angles of dust emission to inform future surveys and average the results over larger areas to remove dust polarization from low-resolution inflationary gravitational wave experiments \cite{poh/dodelson:2017,tassis/pavlidou:2015}.

% galaxy evolution

\textbf{Tracing galaxy evolution from the first billion years to Cosmic Noon with the cosmic infrared background} \cite{geach19}: \ccatp will directly trace the evolution of dust-obscured star formation in galaxies since the epoch of galaxy assembly, starting $>$ 10\, billion years ago \cite{riechers13c,geach19}. By combining our CCAT-prime surveys with synergistic work in the optical/near-IR (e.g. LSST, DES, Euclid and WFIRST) we will identify key parameters that regulate star formation (such as environment and matter content) over cosmic time.

Half of the starlight emitted through cosmic time is obscured by dust which absorbs, and re-emits the light that we measure as the cosmic infrared background (CIB) \cite{Burgarella13,Dole06}.  Therefore, to understand the star formation history of the universe, one must measure it both directly through the optical emission from stars that has been redshifted into the near-IR bands, and indirectly from the dust emission redshifted into the submillimeter bands.  At redshifts $<$ 1.7 most of the CIB was resolved into individual galaxies by Spitzer and Herschel, yet at earlier times the fraction drops to just 10\% \cite{Bethermin12}.  CCAT-prime will probe a factor of 2 to 10 times deeper than the confusion limited Herschel surveys, detecting hundreds of thousands of galaxies to redshifts approaching 7, and unveil about 40\% of the star formation at these epochs. The wide-field ($>$ 400 deg$^2$)  CCAT-prime surveys sample large-scale environments that cannot be mapped with ALMA, and down to luminosities that are inaccessible to Herschel. CCAT-prime surveys at 350 $\mu$m are critical to extracting infrared luminosities and dust-obscured star formation rates at far greater precision than possible with longer wavelength (e.g.  850 $\mu$m) surveys alone. 
% local star formation

\textbf{The star formation-ISM cycle in local galaxies with high spectral
resolution submillimeter observations} \cite{bolatto19,simon19,stanke19}: 
\ccatp will investigate the life cycle of stars from cloud assembly through the formation of stars and expulsion of material at stellar death feeding back into the next cycle, specifically addressing such questions as: How does galactic environment affect star formation and determine its location (e.g. filaments, galactic arms, nuclear rings), intensity, and efficiency?  What is the life-cycle of molecular clouds - the birthplace of stars?  Does turbulence trigger, or inhibit star formation?

High spatial resolution over very large fields thereby linking local star formation sites to global parameters are needed to answer these questions. Using the multi-beam heterodyne receiver (CHAI) on CCAT-prime we can reveal structures in the galaxy at size-scales from 0.01 pc to over 4 kpc - a factor of 4$\times$ 10$^{5}$ in scale. CHAI provides the high spectral resolution necessary to trace inflows, outflows and turbulence, and the velocity information necessary to further distinguish among spatially unresolved structures.  Surveys will focus on the submillimeter [CI] lines to trace the CO dark molecular gas, and mid-J CO lines to trace the warm, active molecular gas in regions of turbulent dissipation and feedback in the Milky Way and nearby star forming galaxies. The [NII] 205 $\mu$m line will be used to trace the ionized gas associated with newly formed early-type stars.

\textbf{A new submillimeter window into time domain astrophysics:} 
\ccatp will discover galactic and extragalactic transients at millimeter and submillimeter wavelengths in a survey covering half the sky, including observations of protostar variability (e.g. \cite{yoo/etal:2017,herczeg/etal:2017}), gamma-ray burst after glows (e.g. \cite{whitehorn/etal:2016}), and potentially new solar system objects (e.g. \cite{cowan/etal:2016}).

\clearpage
\textbf{\it \large Second generation science:}

One of the most compelling research areas in millimeter and  submillimeter astrophysics today is using CMB anisotropy measurements to constrain cosmology.  The cosmological science enabled by these measurements is well summarized in the CMB-S4 Science Book \cite{abazajian/etal:2016} and several decadal white papers, including \cite{astro2020Battaglia,astro2020Green,astro2020Kallosh,astro2020Mantz,astro2020Meerburg}.  These measurements would also enable great progress in many of the ``first generation'' science areas described above.  The \ccatp telescope and site are uniquely capable of advancing the CMB-S4 legacy survey science cases by deploying a second generation ``CMB-S4-scale" receiver on \ccatp and using it in collaboration with the CMB-S4 and/or Simons Observatory projects to observe the majority of the sky at millimeter and submillimeter wavelengths.

The first generation \ccatp measurements will contribute to improved characterization of secondary CMB anisotropies, including the SZ effects, gravitational lensing, and Rayleigh scattering.  These measurements will inform the second generation \ccatp CMB science optimization, which is currently being developed with input from the CMB-S4 and Simons Observatory teams.  This collaborative approach has evolved naturally given that the \ccatp telescope design was adopted for the Simons Observatory Large Aperture Telescope (SO-LAT) and as the reference design for CMB-S4.  As a result, both CMB-S4 and Simons Observatory researchers have at times included submillimeter capabilities being considered for \ccatp in science forecasts.  For example, CMB-S4 galaxy cluster forecasts clearly demonstrate the added value of \ccatp submillimeter measurements for extracting cluster properties \cite{HillPerCom2019,mittal/etal:2017}.  
CCAT-prime will provide unique contributions to the search for primordial gravitational waves by making submillimeter measurements of the CIB that can be used to delens the CMB, thereby reducing the effects of gravitational lensing that act as a source of confusion for primordial gravitational wave searches
measurements 
(e.g. \cite{sherwin/schmittfull:2015,manzotti/etal:2017}). The complementarity of CIB-based and internal reconstruction methods for CMB delensing
will enable valuable cross-checks on both approaches, and depending on the survey areas and complexity of delensing systematics, CIB delensing could potentially prove to be more valuable for constraining inflation. 

\ccatp is also an ideal platform for wide-field, submillimeter line intensity mapping surveys, as its first generation measurements will demonstrate. Improvements in mapping speed by 1-2 orders of magnitude can be achieved by fully populating a next generation cryostat or through switching to spectometer-on-chip technologies, if they reach maturity in the next 5 years. This would shed light on a wide array of puzzles in
galaxy evolution, cosmology, and fundamental physics (e.g., \cite{kovetz19}).

Time-domain measurements, such as searches for additional planets in our solar system \cite{cowan/etal:2016}, measurements of gamma-ray-burst afterglows \cite{whitehorn/etal:2016}, and stellar variability studies \cite{yoo/etal:2017} have all been shown to benefit from both the shorter wavelength and improved resolution of CCAT-prime submillimeter bands.  The high elevation site and improved surface accuracy of \ccatp also provide greater sensitivity for CMB measurements at millimeter wavelengths needed for the core science goals of CMB-S4 and the Simons Observatory.  This combination of new science capabilities and more sensitive  CMB measurements make development of a large camera that utilizes the entire \ccatp field of view a high priority this decade.

\clearpage

\textbf{\large Technical Overview}

\begin{figure}[h]
\begin{center}
\includegraphics[width=6in]{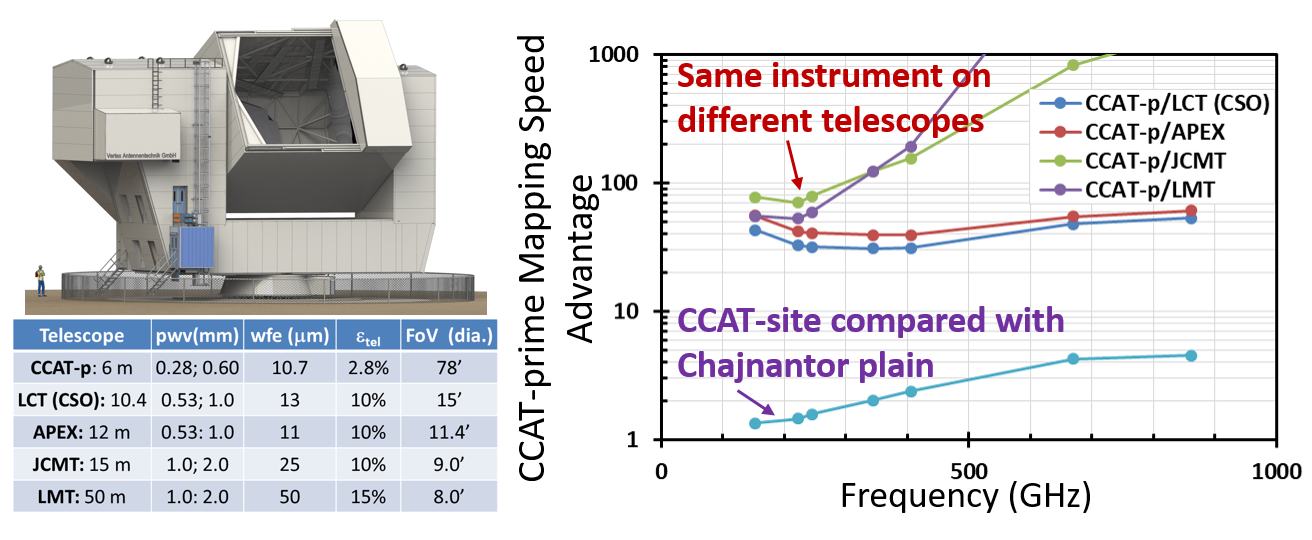}
\end{center}
\vspace{-0.2in}
\caption{\small 
{\bf Site and telescope summary.}  
{\it Top Left:} The CCAT-prime telescope is being built by Vertex Antennentechnik.
{\it Right:} Continuum mapping speed comparisons. The top four lines (green \cite{holland/etal:1999}, purple \cite{wilson/etal:2008}, red \cite{APEX:2006}, and blue \cite{CSO:1994}) are mapping speed ratios at $45^{\circ}$ elevation for a camera with 1.5 $\lambda$/D spaced horns that fills the FoV of a single instrument module in CCAT-prime compared with an identical camera on other coeval telescopes, including the proposed Leighton Chajnantor Telescope (LTC, formerly the CSO) relocation to Chile. We make the comparison for intensity mapping referred to a 1' beam for the top two weather quartiles (PWV \cite{Radford:2016}, wave-front error (wfe), emissivity $\epsilon_{tel}$, and field-of-view (FoV) assumptions at left). CCAT-prime is \textgreater 20$\times$ faster in all cases \textit{per instrument module}. 
Up to 7 instrument modules like this could be deployed in \pcam. The lower, aqua-colored line shows the mapping speed advantage gain per pixel for the CCAT-site on Cerro Chajnantor over the Chajnantor plain \cite{Radford:2016}. The similar SO telescope is only designed to observe at $<$300\,GHz ($\sim$3 left points on curve). 
}
\label{fig:telescope}
\vspace{-0.05in}
\end{figure}

The enabling parameters for \ccatp are driven by the telescope and site.  The 6-m, off-axis, cross-Dragone submillimeter telescope \cite{niemack:2016} delivers an extraordinarily wide diffraction-limited field of view (FoV): from 2\deg\ in diameter at 860 GHz (0.35\,mm) to 8\deg\ at 100 GHz (3\,mm).  In addition, the observatory will be located at 5600-m on Cerro Chajnantor in the Atacama Desert of Northern Chile inside Parque Astron\'{o}mico de Atacama \cite{bustos/etal:2014} above the ALMA Observatory, resulting in improved atmospheric transparency \cite{Radford:2016}.  
The Observatory will have two first generation instruments, \pcam and CHAI.  \pcam will be allocated 75\% of the available observing time over the first five years (over 12,000 hours).  A second instrument, the CCAT Heterodyne Array Instrument (CHAI) is being constructed by the University of Cologne to perform velocity resolved spectral line observations of the Milky Way and nearby galaxies and will use the remaining 25\% of the time.  Figure \ref{fig:telescope} shows the nominal performance gain of CCAT-prime relative to other observatories.  Continuum survey sensitivity is given in Table \ref{tab:surveys}, with more details in  \cite{choi/etal:2019}.

\begin{figure}[h!]
 \noindent
 \begin{center}
 \includegraphics[height=2.2in]{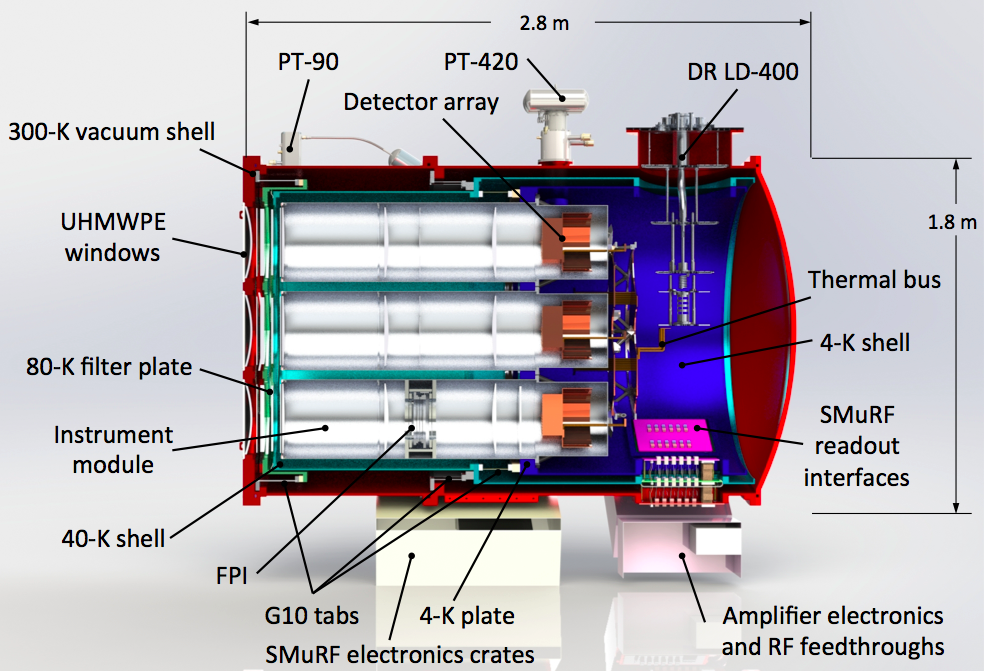}
 \includegraphics[height=2.2in]{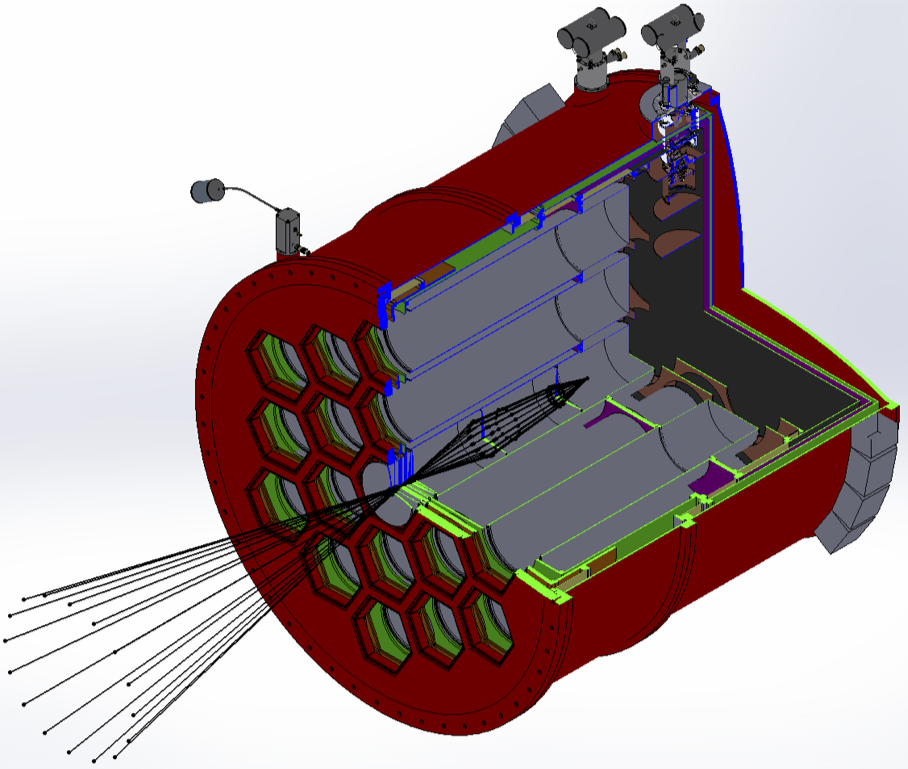}
 \end{center}
 \caption{\small {\it Left:} Cutaway of the \pcam cryostat, which is capable of housing seven instrument modules \cite{vavagiakis/etal:2018}. {\it Right:} Preliminary design for the second generation nineteen module \ccatp instrument. Note that this cryostat is considerably larger than \pcam, while the instrument modules inside both cryostats are the same size.  This model was developed in parallel with the Simons Observatory thirteen module cryostat \cite{zhu/etal:2018} and has recently been adopted as the CMB-S4 reference design \cite{cmb-s4-ref-design:2019}.
 }
 \label{fig:cameras}
 \end{figure}

\pcam is a modular instrument capable of supporting up to seven independent instrument modules to pursue complementary science goals (\cite{stacey/etal:2018}, Figure~\ref{fig:cameras}). These seven modules, or optics tubes, will support broadband measurements at five different frequencies (220, 270, 350, 410, 860 GHz) in parallel with spectroscopic measurements (with $R=100$ between 210 -- 420 GHz) to enable the science goals described above.

The instrument module design details are presented in several recent conference proceedings, including \cite{vavagiakis/etal:2018, dicker/etal:2018, zhu/etal:2018, gallardo/etal:2018}. The nominal module design is being developed collaboratively with the Simons Observatory, but critical changes to the module designs are needed to achieve the \ccatp science goals. Each broadband polarization-sensitive module will include filters, lenses, and three 15\,cm diameter detector arrays.  The \ccatp spectrometer modules require low-temperature Fabry-Perot Interferometer optics that will be installed at the Lyot stop in the relevant modules. For millimeter wavelength detectors the most mature and compelling detector technology is superconducting transition-edge-sensor (TES) bolometers read out by SQUIDs \cite{vavagiakis/etal:2017, crowley/etal:2018}.  
However, it appears that readout challenges will limit development of TES arrays with more than $\sim$2000 detectors per 15\,cm array (e.g., \cite{cmb-s4-technology-book}), which is significantly fewer than the optimal number of detectors at submillimeter wavelengths (and even up to $\sim$1.4\,mm) on \ccatp .  
The most promising detector technology for these submillimeter wavelengths is kinetic inductance detectors, which offer the potential for greatly simplified readout (see below) \cite{Day2003, hubmayr/etal:2015}.

The second generation \ccatp instrument follows the CMB-S4 reference design, which is similar to \pcam but includes nineteen instrument modules instead of only seven (Figure~\ref{fig:cameras}).  This combined with the deployment of more detector arrays per module (as described in the CMB-S4 decadal survey report) will boost the mapping speed of \ccatp by roughly a factor of four compared to a fully populated \pcam.

{\bf Technology Drivers:}

Several of the technology drivers for the measurements described above are being addressed and optimized by a combination of efforts from \ccatp, SO, and CMB-S4.  

Both \ccatp and Simons Observatory are building and will soon be deploying instrument modules (also known as optics tubes) that will serve as prototypes for a CMB-S4-scale project.  When observations begin, these modules will be the largest deployment of superconducting TES detectors yet. The CMB-S4 project is also supporting research and development that is important for scaling these projects to CMB-S4 sensitivity, systematics, and instrument production levels. 

As described above, current TES detector array technologies that can read out two kilopixels per detector wafer \cite{henderson/etal:2016b,cmb-s4-technology-book} are well matched to the CCAT/SO instrument module designs for wavelengths longer than $\sim$1.4\,mm. At shorter submillimeter wavelengths the sensitivity can be improved considerably by increasing the number of detectors per wafer.  The most promising approach for this in the coming decade is using KIDs instead of TESes due to the much simpler readout architecture. KIDs are being used successfully for astrophysical measurements in the NIKA-2 instrument on IRAM \cite{vanRantwijk2016}, and will soon be deployed on both the BLAST-TNG balloon observatory \cite{dober/etal:2016} and the TolTEC camera for the LMT \cite{austermann/etal:2018}.  However, none of those projects will reach the required readout scale to take full advantage of \pcam on \ccatp. 

Breakthroughs in submillimeter astrophysics in the coming decade will require advances in submillimeter detector arrays and readout. \pcam on \ccatp is the ideal platform to enable these advances by deploying submillimeter detector arrays with $\sim10^5$ detectors and using them for novel measurements.  KID detector arrays are rapidly exceeding the detector densities of TES arrays, such as the TolTEC 270\,GHz array with over 4000 KIDs per wafer, 
compared to 2000 TESes on the highest density wafers. For \ccatp at first light we plan to deploy KID arrays operating at both 270\,GHz and 350\,GHz that will each have roughly 4000 and 6000 KIDs, respectively.  Scaling these arrays to all five \pcam bands from 220\,GHz to 860\,GHz with the same $F\lambda$ pixel spacing would result in having between 3,000 and 40,000 KIDs per wafer.  A fully populated \pcam instrument could illuminate 3 wafers each at 220, 270, 350, 410, and 860\,GHz in addition to 6 spectrometer wafers, which would sum to roughly 200,000 detectors!  

Deploying detector arrays at this scale is a high priority for submillimeter astrophysics, although, the readout technologies required for this need additional development. Fortunately, industry is driving advances in microwave systems. For example, the new Xilinx software-defined radio RFSoC system\footnote{Xilinx RFSoC website: \href{https://www.xilinx.com/products/silicon-devices/soc/rfsoc.html}{https://www.xilinx.com/products/silicon-devices/soc/rfsoc.html}} includes the hardware to meet the needs of next generation KID readout.  Specifially, each RFSoC has sufficient bandwidth to acquire data from roughly 5000 KIDs, which is almost ten times more than existing KID readout systems (such as the ROACH-2).  In other words, approximately 40 Xilinx RFSoC boards would be needed to readout a fully populated \pcam, versus, approximately 400 ROACH-2 systems.  The RFSoCs are not yet ready for use due to the need to develop KID-optimized RFSoC firmware, which can then be tested and deployed on astronomical observatories like \ccatp. Other high bandwidth readout systems are also being developed, such as the SLAC Microresonator Radio Frequency (SMuRF) electronics\cite{henderson/etal:2018}, and different approaches to solving this readout challenge are being explored.  In addition to submillimeter astrophysics, this readout development research is expected to benefit observatories operating at many other wavelengths spanning from optical to radio.  \ccatp is an ideal telescope for the initial deployment of these technologies because of the high-throughput, modularity, relatively low operating costs, and flexibility of the team. 

The combination of the ongoing instrument module development with support for development of submillimeter KID arrays and readout systems for \pcam will enable the technology advances needed for both CMB-S4-scale instruments and next generation submillimeter astrophysics.

\clearpage                   

\begin{figure}[h!]
\vspace{-0.1in}
\noindent
\begin{minipage}{.3\textwidth}
\caption{\small 
CCAT organization chart showing members of the German and Canadian (CATC) consortia.  Other participants in Canada include St. Mary's University, University of Manitoba, University of Lethbridge and University of Alberta.  The Board of Directors has representatives from the these two consortia and Cornell.
}
\label{fig:orgchart}
\end{minipage}%
\begin{minipage}{.7\textwidth}
\begin{center}
\includegraphics[width=3.5in]{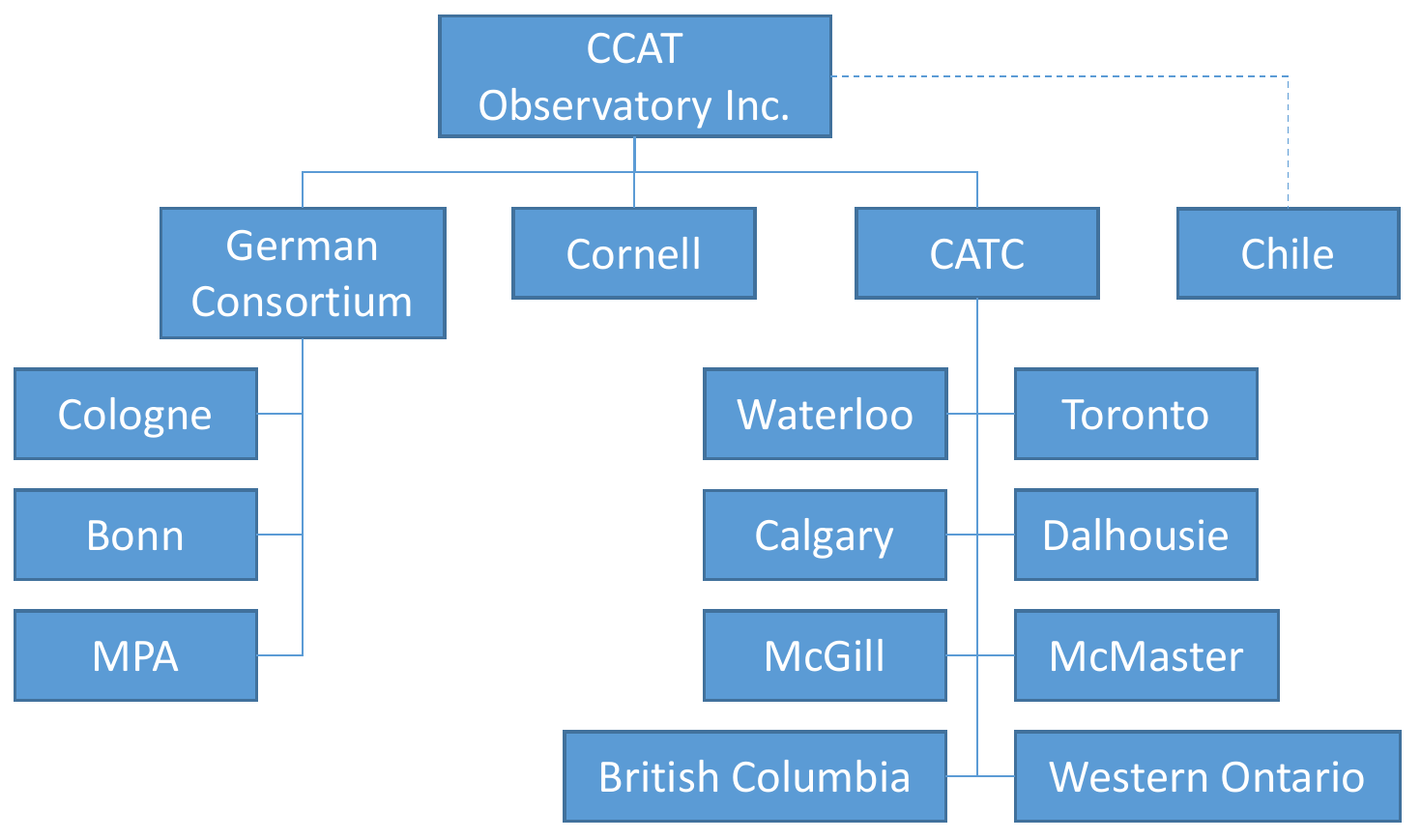}
\end{center}
\end{minipage}%
\end{figure}

\textbf{Organization, Partnerships, and Current Status:}
\ccatp is being built by CCAT Corporation which is independently funded by an international consortium consisting of Cornell University, a German partnership led by the University of Cologne, and a Canadian partnership led by the University of Waterloo. Figure \ref{fig:orgchart} shows organizational chart and member institutions, as well as Chilean participation. The Tokyo Atacama Observatory (TAO) shares the mountain top.  CCAT has a draft agreement to share common costs such as road maintenance, power, etc.  With funding from NSF (or other public sources), CCAT-prime will bring in community members to participate in survey planning and provide general access to curated data sets.  

\begin{figure}[h]
\begin{center}
\includegraphics[width=6in]{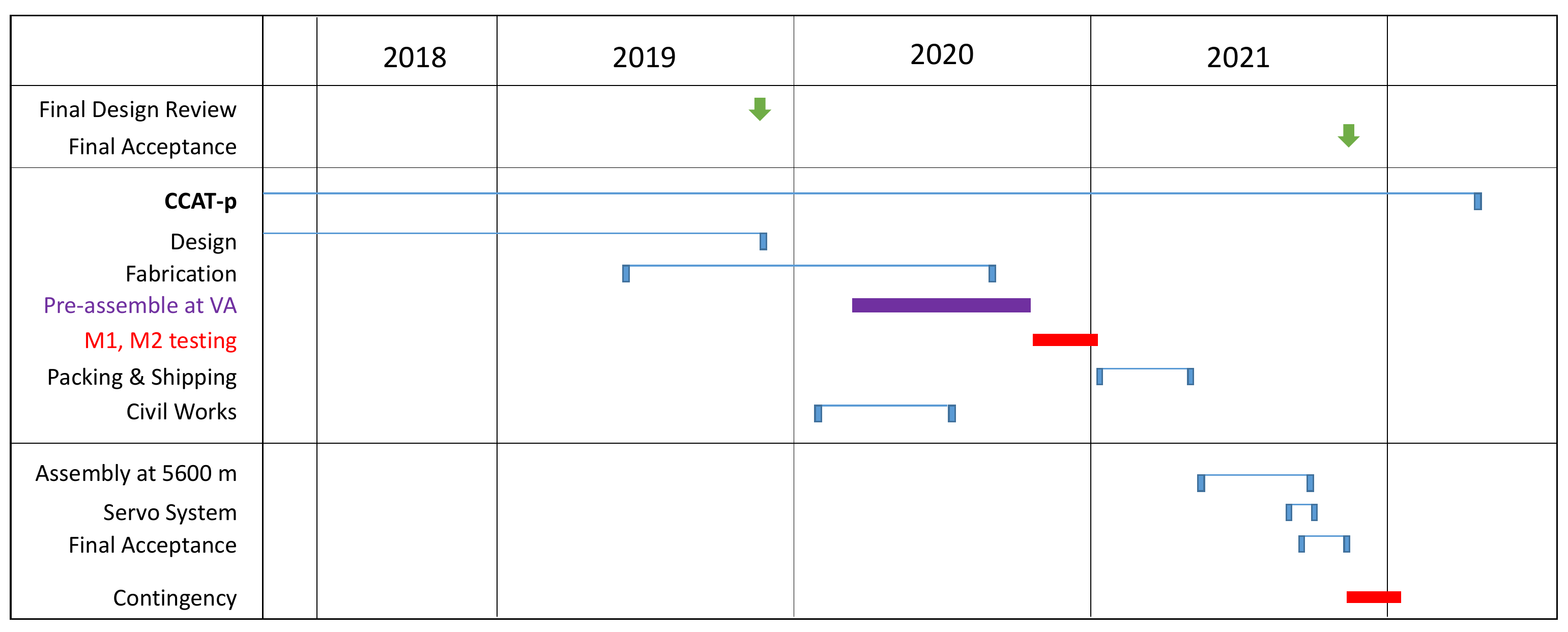}
\end{center}
\vspace{-0.2in}
\caption{\small 
Top-level CCAT-prime observatory schedule showing major milestones.  First light is scheduled for the last quarter of 2021. The initial science surveys occupy the first five years of operations.  The observatory would then nominally transition to second generation science.  The lifetime of the observatory is expected to be 10-20 years.
}
\label{fig:schedule}
\vspace{-0.05in}
\end{figure}

\textbf{Schedule:}  
The telescope is being built by  Vertex Antennentechnik GmbH under a fixed-price contract to the CCAT Corporation.  An in situ laser interferometric metrology system developed by Etalon-AG (Germany) will be used to align the mirror panels to a few microns. Site preparation and development (civil works) are performed in parallel with the telescope construction.  Figure \ref{fig:schedule} shows the schedule.  Fabrication of long lead items for the telescope is underway.  To ensure efficient and safe assembly on the mountain, the telescope is fully assembled and aligned in Europe, and then partly disassembled for shipment to Chile.  Following re-assembly on the mountain, the surface accuracy will be verified using the laser system and tower holography.     

\vspace{-0.1in}

\clearpage                 

\textbf{Cost Estimates:}
Table \ref{tab:funding} gives the run out costs in real year dollars for CCAT-prime.  These estimates are updated quarterly. The basis of costs for the Observatory are a fixed-price contract for the telescope and bids for the infrastructure work.  The facility operations costs are based on a bottoms-up estimate (fuel, road maintenance, site fees, on-going maintenance, etc.).  Science operations are based on sizing of the team to handle data reduction and analysis, curation and maintenance of the archive for the community, support of workshops for the community, etc..

Also included in Table \ref{tab:funding} are funding commitments from private and international sources and aspirational funding from federal sources.  Although the construction of CCAT-prime is covered by private and international funding,
public funding is needed to complete the Prime-Cam instrument and support operations.  The total funding line includes five years of operations. 

CCAT-prime is ideally suited to participate in the next generation of CMB Stage-4 scale science for which  the costs will be of order \$30 - 50M for instrumentation and operations.

\begin{table}[h!]
\vspace{-0.2in}
{\small
\hfill{}
\small
\begin{center}
\vspace{-.2in}
\caption{Costs and Contributions}
\vspace{0.03in}
\begin{tabular}{|c| c| c c c| c |}
\hline
 &  & \multicolumn{3}{c|}{Contribution} &  \\
 \hline
Item       & Cost             & Federal$^a$      & International  & Private & Comments \\
         &    (M\$)  & (M\$) & (M\$) & (M\$)  & \\
\hline
Observatory     & 26.5$^b$    & -  & 10 & 17.35   &  includes telescope and infrastructure \\
Prime-Cam$^c$   & 13    & 8  & 2  & 3      &  first-light instrument \\
CHAI$^d$        & 9     & -  & 9  & -       &  first-light instrument \\
\hline
Facility Ops/yr & 2.5   & 1  & 1  & 0.5     &  \\
Science Ops/yr  & 2     & 1  & 1  & -       &  \\
\hline
\hline
Total           & 71.5  & 18 & 31 & 22.85 & through 5 years of operations\\
\hline
\end{tabular}
\label{tab:funding}
\end{center}}
{\small { 
$^a$\,Aspirational funding levels for instrument development and operations.
$^b$\,Excludes cost of constructing power plant on plateau.
$^c$\,Prime-Cam can support up to 7 science modules at a cost of approximately \$2M per modules based on currently technologies.  The baseline is 2 modules at first-light with 3 more added within 2 years.  
$^d$\,Fully funded by German consortium. 
}}
\vspace{-.2in}
\end{table}

\noindent\makebox[\linewidth]{\rule{\textwidth}{1pt}} 
\vspace{-0.2in}
\begin{table}[h!]
\vspace{-0.2in}
{\small
\hfill{}
\small
\begin{center}
\vspace{-.2in}
\caption{Overview of baseline Prime-Cam survey performance$^c$ \cite{stacey/etal:2018,choi/etal:2019} }
\begin{tabular}{*7c}
\hline
Survey       & Field ID             & LST range & Area      & Time & Sensitivity$^c$ & Supporting \\
&          &    [h]  & [deg$^2$] & [hr] & (@ representative $\nu_{\rm obs}$[GHz])    & Surveys$^b$ \\
\hline
EoR$^a$      & E-COSMOS             & 7.0-13.0  & 8   & 2000 & 0.02\,MJy\,sr$^{-1}$\,bin$^{-1}$ @ 220 & 1 \\
             & E-CDFS               & 23.5-7.0  & 8    & 2000 & 0.02\,MJy\,sr$^{-1}$\,bin$^{-1}$ @ 220 & 2 \\
             & HERA-Dark            & 13.0-23.5      & 8    & {\footnotesize\em (filler)} & 0.02\,MJy\,sr$^{-1}$\,bin$^{-1}$ @ 220 & 3 \\
DSFG         & Stripe 82            & 20.0-5.5  & 300   & 500 & 2.5\,mJy\,beam$^{-1}$ @ 860 & 4 \\
& GAMA9/12/15 & 5.5-20.0  & 110    & 180  & 2.5\,mJy\,beam$^{-1}$ @ 860 & 5 \\
SZ/CMB   & AdvACT/SO            & all       & 12,000 & 4000 & 11\,$\mu$K/arcmin$^{2}$ (CMB) @ 270 & 6 \\
\hline
\label{tab:surveys}
\end{tabular}
\vspace{-.25in}
\end{center}}
{\small { 
$^a$Spectroscopy; sensitivities provided for $R$=100.
$^b$(1) Deep Subaru HSC+PSF spectroscopy \& COSMOS X-Ray-to-meter-wave multiwavelength survey; (2) deep Euclid grism spectroscopy (upcoming), HERA HI 21\,cm (upcoming), \& H-UDF/CDF-S multiwavelength surveys (incl.\ JWST GTO); 
(3) HERA HI 21\,cm (upcoming), VLASS; 
(4) SDSS, HeLMS/HeRS Herschel/SPIRE, VLASS; 
(5) GAMA, H-ATLAS Herschel/SPIRE, ACT, VLASS; 
(6) Planck, SDSS, DES, ACT, SO, DESI, LSST, eROSITA (upcoming).
$^c$Preliminary sensitivity model; a more advanced model is now available in Choi et al. (2019) \cite{choi/etal:2019}. 
}}
\vspace{-.2in}
\end{table}

\clearpage           
\smallskip
\bibliographystyle{th_custom2}{}
\bibliography{ccat_refs}

\end{document}